\documentclass[english,twocolumn,superscriptaddress,
  longbibliography]{revtex4-1}

\usepackage{xcolor}
\usepackage{graphicx}
\usepackage{amsmath}
\usepackage{amsfonts}
\usepackage{color}
\usepackage{babel}
\usepackage{url}
\usepackage{hyperref}
\usepackage{tikz}
\usepackage{CJKutf8}
\usepackage[title]{appendix}

\definecolor{lime}{HTML}{A6CE39}
\DeclareRobustCommand{\orcidicon}{
	\begin{tikzpicture}
	\draw[lime, fill=lime] (0,0) 
	circle [radius=0.16] 
	node[white] {{\fontfamily{qag}\selectfont \tiny ID}};
	\draw[white, fill=white] (-0.0625,0.095) 
	circle [radius=0.007];
	\end{tikzpicture}
	\hspace{-2mm}
}

\foreach \x in {A, ..., Z}{\expandafter\xdef\csname
  orcid\x\endcsname{\noexpand\href{https://orcid.org/\csname
      orcidauthor\x\endcsname} {\noexpand\orcidicon}} }

\begin{document}

\title{Nonlinear optics from hybrid dispersive orbits}

\author{Yongjun Li\orcidA{}}\thanks{email: yli@bnl.gov}
\affiliation{Brookhaven National Laboratory, Upton 11973, New York, USA}

\author{Derong Xu} \affiliation{Brookhaven National Laboratory, Upton
  11973, New York, USA}

\author{Victor Smaluk} \affiliation{Brookhaven National Laboratory, Upton
  11973, New York, USA}

\author{Robert Rainer\orcidB{}} \affiliation{Brookhaven National
  Laboratory, Upton 11973, New York, USA}
 
\begin{abstract}
  In this paper we present an expansion of the technique of characterizing
  nonlinear optics from off-energy orbits (NOECO)~\cite{olsson2020} to
  cover harmonic sextupoles in storage rings. The existing NOECO technique
  has been successfully used to correct the chromatic sextupole errors on
  the MAX-IV machine, however, it doesn't account for harmonic sextupoles,
  which are widely used on many other machines. Through generating
  vertical dispersion with chromatic skew quadrupoles, a measurable
  dependency of nonlinear optics on harmonic sextupoles can be observed
  from hybrid horizontal and vertical dispersive orbits. Proof of concept
  of our expanded technique was accomplished by simulations and beam
  measurements on the National Synchrotron Light Source II (NSLS-II)
  storage ring.
\end{abstract}
 
\maketitle

\section{\label{sect:intro}introduction}

  Characterizing the nonlinear optics of storage rings is becoming more
  essential with the introduction of higher order multipole magnets in
  accelerator design. Errors from the higher order multipoles have been
  observed to degrade machine performance, such as reduction of dynamic
  aperture, energy acceptance, etc. Some efforts have been made to
  identify the nonlinear multipole errors by measuring distorted resonance
  driving terms~\cite{franchi2014}, which requires a complicated
  Hamiltonian dynamics analysis. A more practical technique for measuring
  the nonlinear optics from off-energy closed orbits (NOECO) was reported
  and demonstrated on the MAX-IV ring~\cite{olsson2020}. Significant
  improvements on its dynamic aperture and beam lifetime were observed
  after correcting sextupole errors. Desired results were obtained while
  testing the NOECO technique on the ESRF-EBS ring as
  well~\cite{liuzzo2022}. However, the dependency of nonlinear optics on
  off-energy orbits is only measurable for chromatic sextupoles. The
  horizontal dispersion seen by chromatic sextupoles are usually quite
  large, as to effectively correct the chromaticity. This technique,
  however, doesn't apply to harmonic sextupoles, which do not see the
  first order linear dispersion. Harmonic sextupoles are used in almost
  every third-generation light source ring, and some fourth-generation
  diffraction-limited machines, such as the ALS-U ring~\cite{steier2019}.
  They are even being used in the design of a future electron-ion collider
  ring~\cite{cai2022}. As such, an expansion of the existing NOECO
  technique to correct for the harmonic sextupoles would be useful due to
  their common, integral use in current and future accelerator design. In
  the National Synchrotron Light Source II (NSLS-II)
  ring~\cite{dierker2007}, the number of harmonic sextupoles are greater
  than the number of chromatic sextupoles (180:90). Therefore, correcting
  harmonic sextupole errors is important for improving machine performance
  due to their greater influence. In this paper, we outline our expansion
  on the capabilities of existing sextupole correction techniques to
  accommodate for the harmonic sextupoles.

  A straightforward method for calibrating harmonic sextupoles for
  correction would be to temporarily convert them to chromatic ones. This
  could be achieved by tuning the quadrupoles inside achromats to generate
  a commensurate amount of dispersion at the locations of harmonic
  sextupoles~\cite{olsson2022}. However, this method would require a
  significant modification of the original linear lattice. Implementations
  during online measurements, such as updating the nonlinear optics
  dependency for different leaked dispersion bumps would also be
  complicated. Another method would be to generate local orbit bumps,
  calibrated through the sextupoles, and then measuring the optics
  distortion with different bump parameters. This method would not only
  require sufficient beam position monitors (BPMs) that neighbor the
  sextupoles, but would also be complicated to implement. In real-world
  applications, it is time-consuming to form perfectly closed local bumps
  with orbit correctors, and then to update the optics dependence on these
  bump settings~\cite{choi2023}. While the above methods would be capable
  of achieving the desired outcome, they are not practical when
  considering the limitations of routine operations of user facilities.

  When a sextupole sees vertical dispersion, the nonlinear optics of the
  off-energy orbits will also depend on its gradient
  $K_2=\frac{1}{(B\rho)_0}\frac{\partial^2B_y}{\partial x^2}$, normalized
  with the beam rigidity $(B\rho)_0$. A vertical dispersive wave can be
  generated through chromatic skew quadrupoles. In most light source
  rings, skew quadrupoles are widely equipped to control the residual
  vertical dispersion and linear coupling. Usually, a considerable amount
  of vertical dispersion can be generated, but only introduces weak
  coupling when the Betatron tune has sufficiently diverged from the
  linear difference/sum resonance. Thus, the nonlinear off-energy optics
  depends on not only chromatic sextupoles, but also on the original
  harmonic ones. In other words, horizontal harmonic sextupoles are
  converted into vertical chromatic ones, which makes their calibration
  and correction possible on hybrid dispersive orbits. In our studies, the
  NSLS-II ring double-bend achromat lattice was used to demonstrate these
  expanded capabilities.

  The remainder of this paper is outlined as follows:
  Sect.~\ref{sect:method} introduces the principle of the technique in
  conjunction with the NSLS-II lattice. In Sect.~\ref{sect:simulation} we
  demonstrate our technique with some simulations. Some beam measurements
  to calibrate both the chromatic and harmonic sextupole errors are given
  in Sect.~\ref{sect:meas}, with the caveat that no real sextupole
  correction can be implemented at this time due to their in-series power
  supplies. Sect.~\ref{sect:hardware} discusses the hardware requirements
  necessary to apply this technique. A brief summary is given in
  Sect.~\ref{sect:summary}.

\section{\label{sect:method}Nonlinear optics on hybrid dispersive orbit}

\begin{figure}[!ht]
    \centering
    \includegraphics[width=\columnwidth]{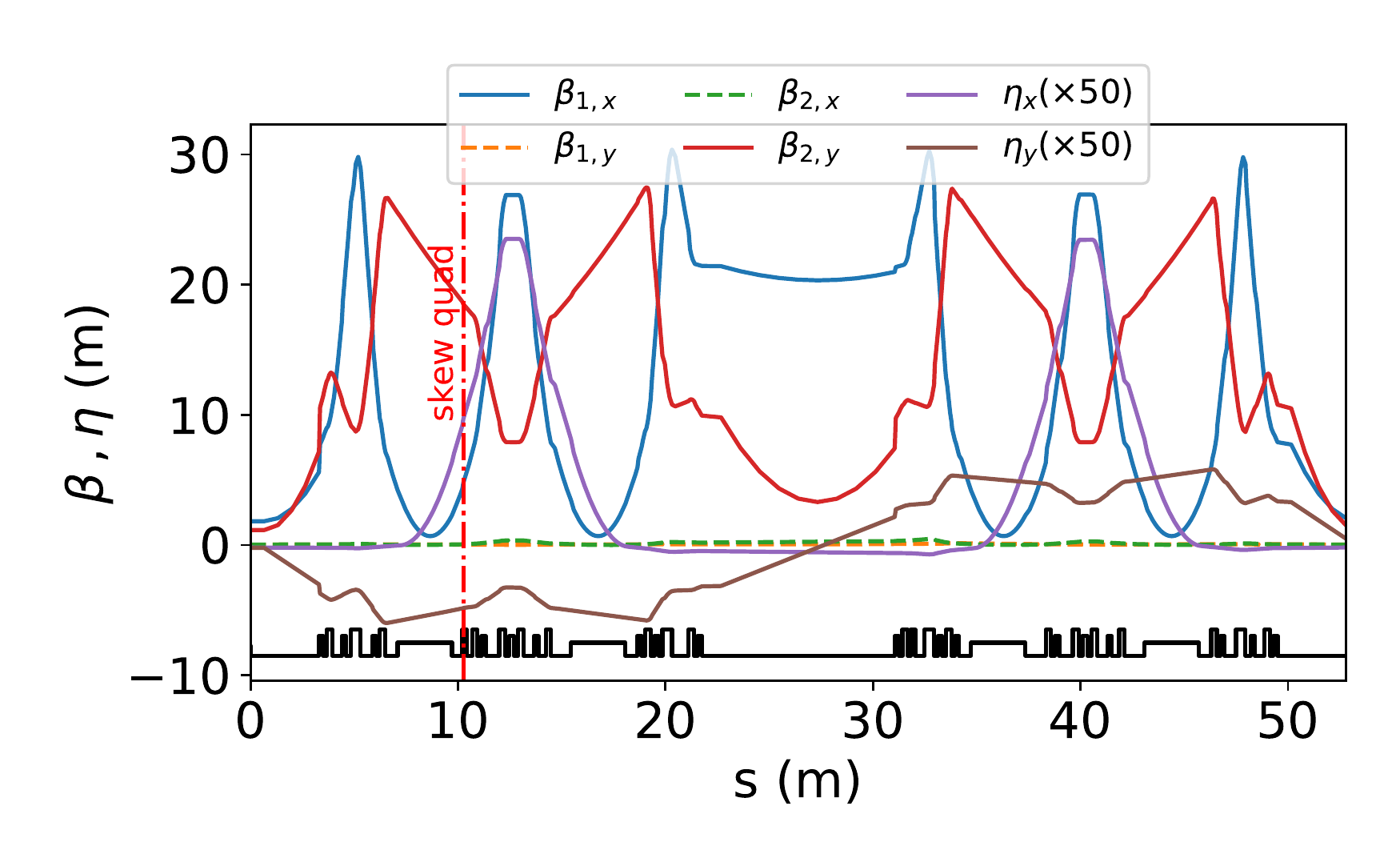}
    \caption{\label{fig:twiss} Exact Ripken Twiss functions for a
      supercell at NSLS-II when a vertical dispersion wave is generated
      with chromatic skew quadrupoles (with a normalized gradient
      $K_{1,s}=\frac{1}{(B\rho)_0}\frac{\partial B_y}{\partial x}=0.070\;
      m^{-2}$). The location of the skew quadrupole is marked with a red
      vertical dash-dot line. Two non-dominated $\beta$-functions (dashed
      lines) indicate that this optics configuration remains weakly
      coupled.}
  \end{figure}

\begin{figure}[!ht]
    \centering
    \includegraphics[width=\columnwidth]{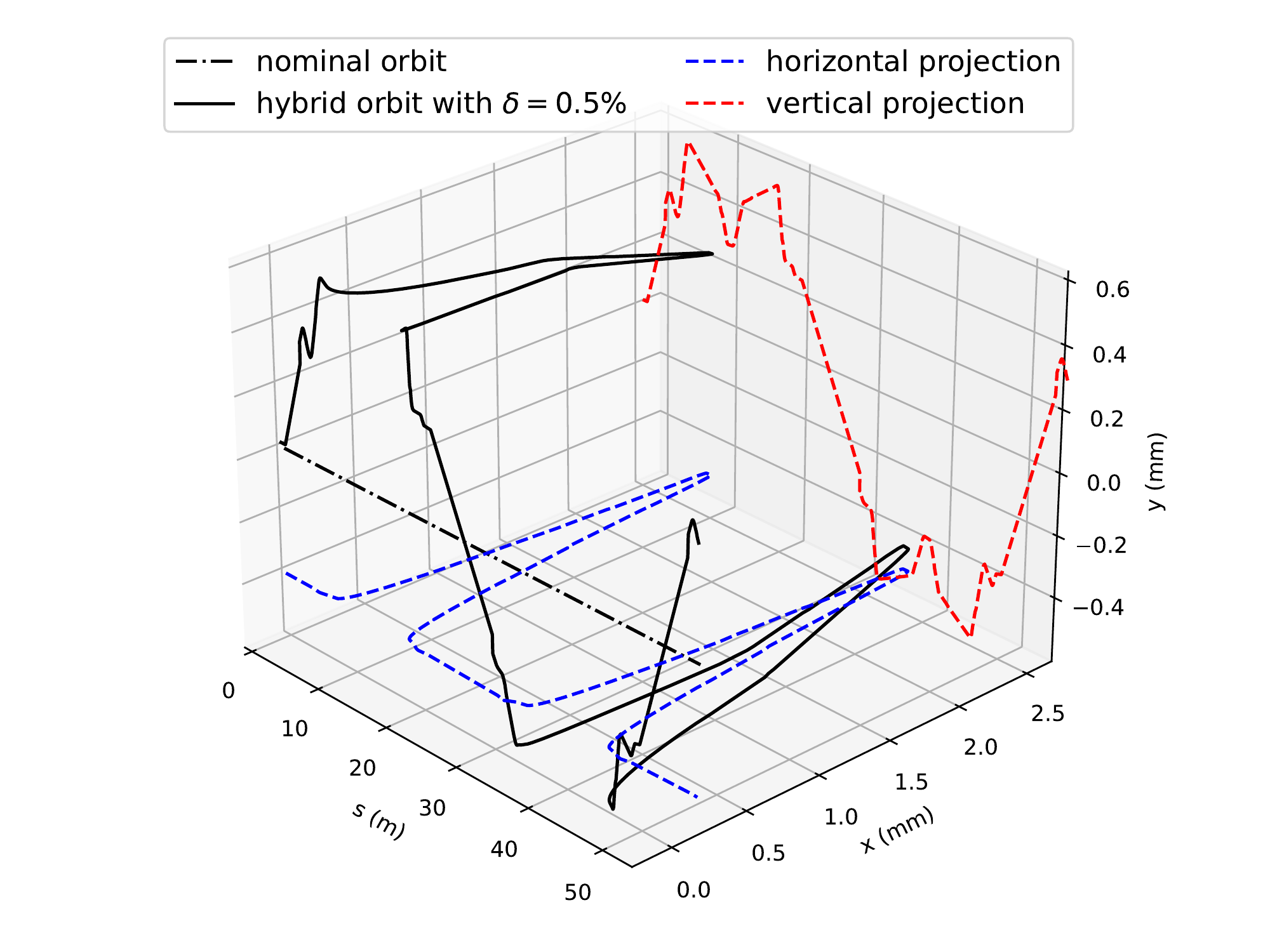}
    \caption{\label{fig:hybridOrbit} Three-dimensional view of the hybrid
      dispersive orbit with $\delta=0.5\%$ for one NSLS-II supercell.}
  \end{figure}

  Chromatic skew quadrupoles (located at horizontally dispersive sections)
  can couple the dispersion function between the horizontal and vertical
  planes. This property is widely used to minimize the vertical beam size
  in most light source rings. At the NSLS-II ring, each odd-numbered cell
  is equipped with one $0.2\;m$ long chromatic skew quadrupole (see
  Fig.~\ref{fig:twiss}). Their maximum gradients are $g_1=0.35\;T\cdot
  m^{-1}$, which is limited by the capacity of their power
  supplies. Assuming we can double their gradients to $g_1=0.70\;T\cdot
  m^{-1}$, a vertical dispersion wave with a $\sim0.1\;m$ amplitude can be
  generated. The necessity for a double gradient will be discussed in
  Sect.~\ref{sect:hardware}. Although these gradients are twice as large
  as the maximum output of their power supplies, they are still quite weak
  compared to other operational quadrupoles with a maximum gradient of
  $g_{1,max}=22\;T\cdot m^{-1}$. Under these conditions, the exact coupled
  optics computed with the Ripken
  parameterization~\cite{borchardt1988,willeke1989} indicates that the
  linear optics remain weakly coupled. In Fig.~\ref{fig:twiss}, the
  non-dominated functions $\beta_{1,y}$ and $\beta_{2,x}$ (dashed lines)
  are observed as very close to zero, while the dominated $\beta_{1,x}$
  and $\beta_{2,y}$ (solid lines) are almost the same as in the uncoupled
  case. The skew quadrupoles also cause a small amount of horizontal
  dispersion to be leaked into the straight sections. Although such small
  residual dispersion could not be solely used to measure the off-energy
  nonlinear optics, its effect is accounted for in our method because the
  exact parameterization has been used. In short, when the machine tune is
  configured to avoid linear coupling resonances, chromatic skew
  quadrupoles can generate a considerable amount of vertical dispersion,
  but only introduce relatively small linear coupling. The newly generated
  vertical dispersion seen by the original harmonic sextupoles can make
  the nonlinear optics on off-energy orbits (as illustrated in
  Fig.~\ref{fig:hybridOrbit}) dependent on their gradients. Therefore,
  this dependence can be utilized for their calibration and correction.

\begin{figure}[!ht]
    \centering
    \includegraphics[width=\columnwidth]{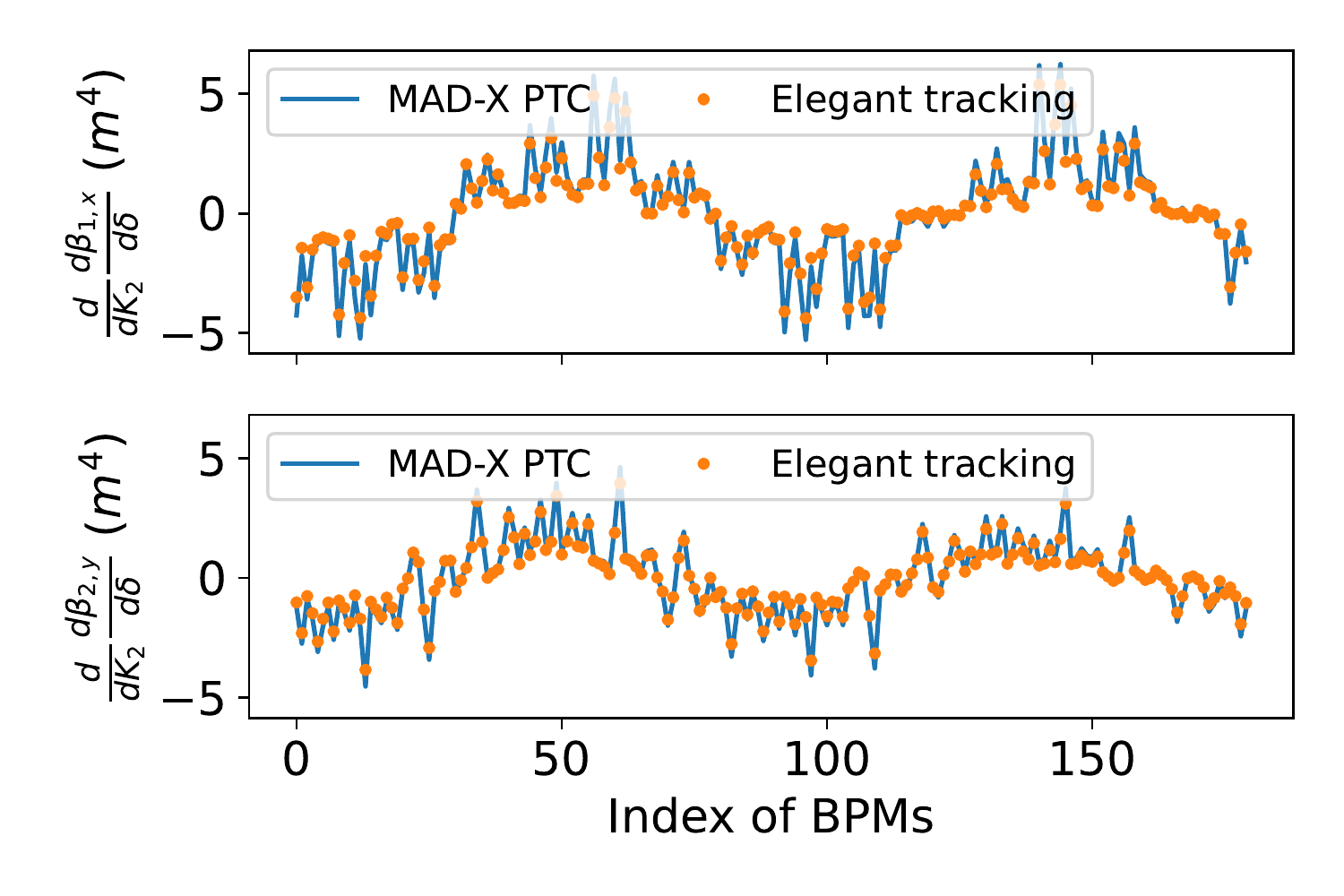}
    \caption{\label{fig:ptc_tracking} Comparison of the optics dependence
      on one harmonic sextupole computed with different methods. Top: the
      response vector in the horizontal plane. Bottom: the response vector
      in the vertical plane}
  \end{figure}

  When horizontal dispersion is seen by sextupoles in an uncoupled linear
  optics configuration, the dependence of the $\beta$-function on the beam
  energy deviation, $\delta=\frac{\Delta P}{P}$, and the normalized
  sextupole gradient, $K_2$, can be formulated~\cite{wiedemann2015}. When
  both horizontal and vertical dispersion can be seen by sextupoles in a
  weakly coupled optics configuration, no such simple formulae are
  available. However, it can be numerically computed with the two
  following methods. Method 1: First, for a given energy offset $\delta$
  and skew quadrupole settings $K_{1,s}$, a hybrid dispersive closed orbit
  can be obtained with iterative tracking. This hybrid dispersive orbit
  now has both the horizontal and vertical offsets. Then a one-turn matrix
  $R$ can be obtained along the dispersive orbit with a truncated power
  series algorithm technique~\cite{berz1988}. From the linear components,
  four coupled Ripken Twiss functions can then be extracted and propagated
  around the whole ring. By slightly tweaking the settings of an arbitrary
  sextupole with a $dK_2$, and repeating the above procedure, the
  dependence of $\frac{d\beta}{d\delta}$ on $K_2$ can be
  determined. Method 2: A direct particle tracking can be implemented with
  the same lattice setting as described in the previous method. When the
  initial particle coordinates are confined within the linear regime, the
  linear one-turn matrix $R$ can also be fitted from turn-by-turn
  trajectories. Then the Ripken Twiss functions can be
  parameterized. After comparing the results for one sextupole using these
  two methods, which were yielded respectively with the \textsc{mad-x} PTC
  analysis~\cite{schmidt2005} and the particle trajectory tracking with
  the code \textsc{elegant}~\cite{borland2000}, the results of both
  methods were consistent~\ref{fig:ptc_tracking}.

\begin{figure}[!ht]
    \centering
    \includegraphics[width=\columnwidth]{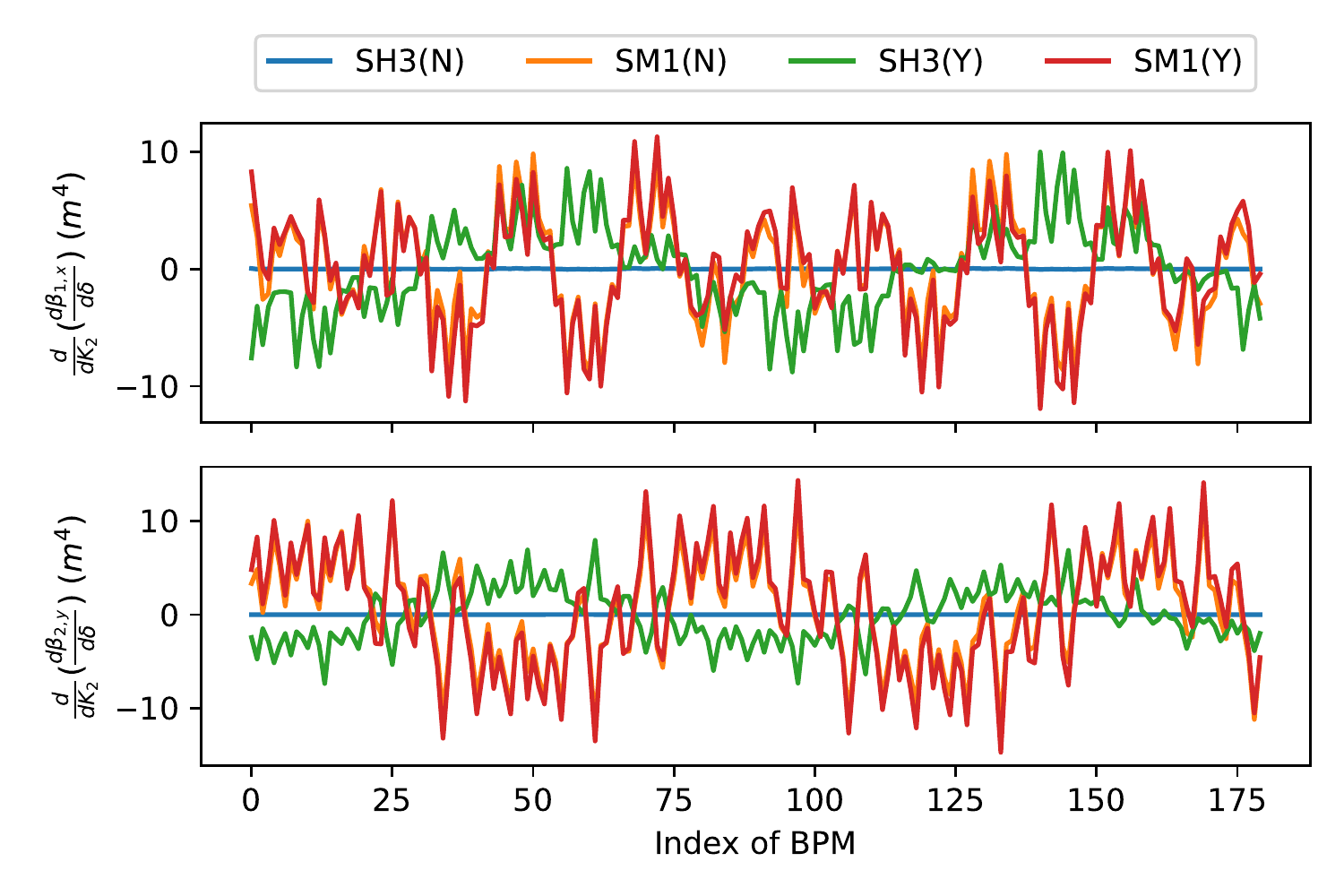}
    \caption{\label{fig:ch_harm} Comparison of the off-energy optics
      dependence on a chromatic sextupole ``SM1'' and a harmonic sextupole
      ``SH3''. On horizontal-only dispersive orbits, the optics dependence
      on the harmonic sextupole (labeled as ``SH3(N)'') is too small to
      measure. On hybrid dispersive orbits, a measurable dependence
      (labeled as ``SH3(Y)'') is observed.}
  \end{figure}

  For demonstration purposes, we choose a harmonic sextupole ``SH3'' and a
  chromatic sextupole ``SM1'' to compute their linear dependence on the
  off-energy optics, i.e., the so-called response vectors, as seen
  below. Only two dominated optics functions
  $\frac{d\beta_{1,x}}{d\delta}$ and $\frac{d\beta_{2,y}}{d\delta}$
  observed at their corresponding BPMs were computed. If no skew quadrupoles are used to
  excite the beam, the optics functions degenerate to the uncoupled
  $\beta_x$ and $\beta_y$. The response vectors computed with and without
  the vertical dispersion are compared in Fig.~\ref{fig:ch_harm}. With
  horizontal-only dispersive orbits, the dependence of off-energy optics
  on ``SH3(N)'' is not measurable in both the horizontal and vertical
  planes. On the hybrid dispersive orbits, a measurable dependence on
  ``SH3(Y)'' can be observed. Note that, for both cases, the dependence of
  the chromatic ``SM1(Y/N)'' is always measurable because it sees a large
  horizontal dispersion. In the meantime, the dependencies are quite
  similar since the optics are only slightly altered.
  
  In principle, chromatic and harmonic sextupole errors can be calibrated
  simultaneously with a sufficiently large vertical dispersion. However,
  chromatic sextupoles usually have stronger responses than harmonic ones,
  particularly when the dispersion in the vertical plane is coupled from
  the horizontal plane. Therefore, for practical purposes, we can uncouple
  the chromatic and harmonic sextupole correction via a two-stage
  correction. Stage-1: correcting chromatic sextupoles first with the
  existing technique~\cite{olsson2020}. Stage-2: generating a vertical
  dispersion wave, then calibrating harmonic sextupoles from hybrid
  dispersive orbits. In this paper, we only focus on the $2^{nd}$ stage
  because the $1^{st}$ stage has already been well studied with the linear
  optics from closed orbits (LOCO) algorithm~\cite{safranek1997} in the
  ref.~\cite{olsson2020}.
  
\begin{figure}[!ht]
    \centering
    \includegraphics[width=\columnwidth]{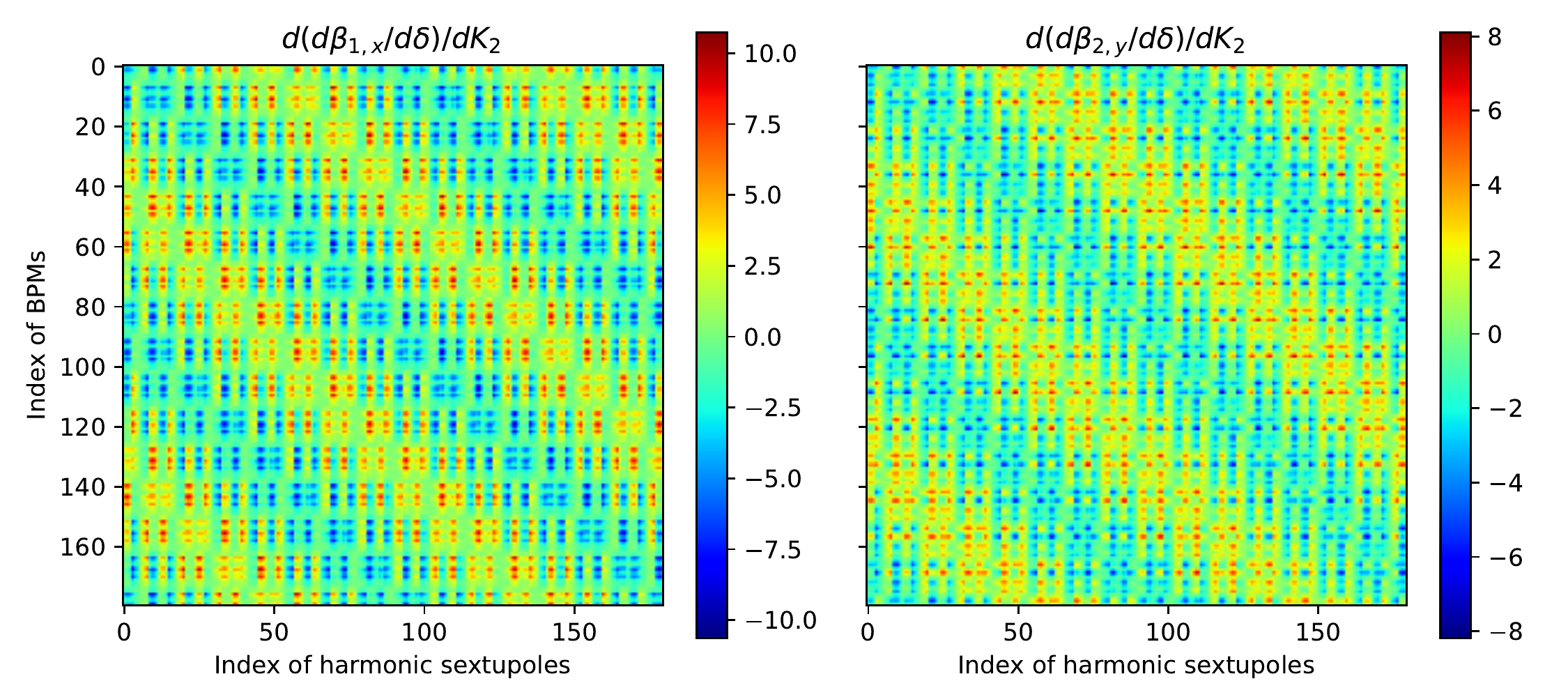}
    \caption{\label{fig:rmxy} Given a hybrid dispersion wave in
      Fig.~\ref{fig:twiss}, the harmonic sextupole response matrices (RM)
      observed at the BPMs are obtained by computing the deviation of
      $\frac{d\beta}{d\delta}$ while slightly changing each individual harmonic
      sextupole from its nominal setting. Left: the dominated
      $\beta_{1,x}$ function's RM in the horizontal plane. Right: the
      dominated $\beta_{2,y}$ function's RM in the vertical plane.}
  \end{figure}

  The $\beta$-functions and phase advances can be measured directly from
  turn-by-turn data using the harmonic analysis~\cite{borer1992}, or the
  numerical analysis of fundamental frequencies (NAFF)
  algorithm~\cite{laskar1992, zisopoulos2013}. Therefore, instead of the
  LOCO algorithm, the dependence of $\frac{d\beta}{d\delta}$ on the
  harmonic sextupole settings $dK_2$, which were deliberately mis-set, was
  used in our simulations. Given a vertical dispersion wave pattern as shown
  in Fig.~\ref{fig:twiss}, the response matrices of
  $\frac{d\beta_{(1,x),(2,y)}}{d\delta}$ dependence on 180 harmonic
  sextupoles were computed with the lattice model and illustrated in
  Fig.~\ref{fig:rmxy}. Here, we only used two dominated
  $\beta_{(1,x),(2,y)}$-functions, and the other two non-dominated ones,
  $\beta_{(1,y),(2,x)}$, were ignored because they are too small to
  measure accurately.
  
\section{\label{sect:simulation}Simulations}
  
  Below we simulated two specific cases to study the performance of our
  expanded sextupole correction technique.

  \subsection{Case 1: two individual isolated errors}

\begin{figure}[!ht]
      \centering \includegraphics[width=\columnwidth]{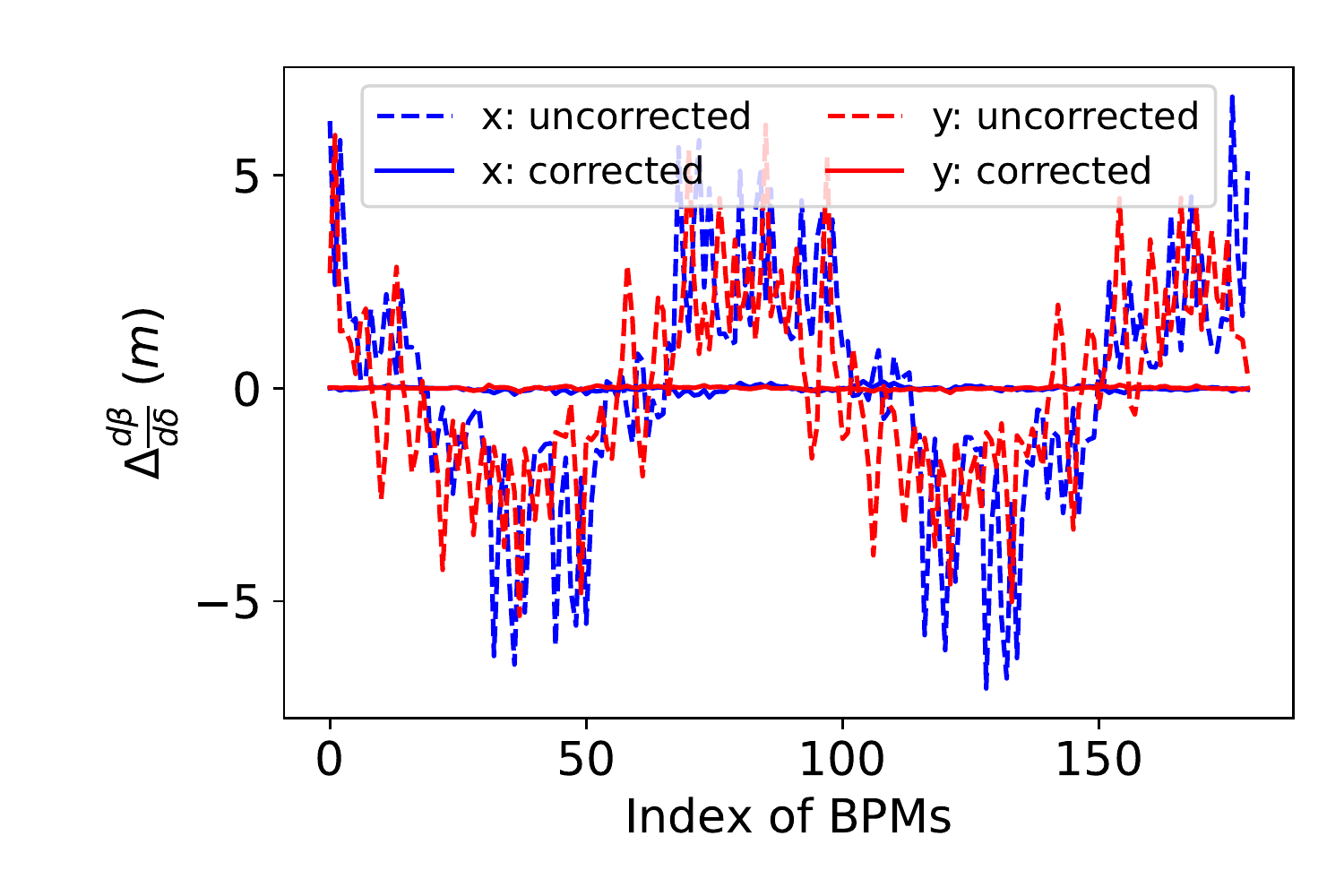}
      \caption{\label{fig:double_distort} Comparison of nonlinear optics
        distortions before and after correction in the presence of two
        isolated sextupole errors. The dashed lines are the uncorrected
        distortions, and the solid lines represent the corrected ones.}
    \end{figure}

    First, we studied a case in which two isolated (far apart from each
    other) sextupoles errors $\Delta K_2=+1.5,\;-1 m^{-3}$ were added onto
    the $32^{th},\;154^{th}$ harmonic sextupoles. The distortions of
    $\Delta\frac{d\beta}{d\delta}$ observed at the BPMs are shown with the
    dashed lines in Fig.~\ref{fig:double_distort}. The needed corrections
    $\Delta K_2$ were obtained by solving the following linear regression
    problem with the response matrices computed in the previous section,
    \begin{equation}\label{eq:correction}
      \Delta\frac{d\beta}{d\delta} \approx M_{x,y}\Delta K_2,
    \end{equation}
    here, $M_{x,y}$ represents a vertically stacked matrix with the
    horizontal and vertical response matrices.

\begin{figure}[!ht]
      \centering \includegraphics[width=\columnwidth]{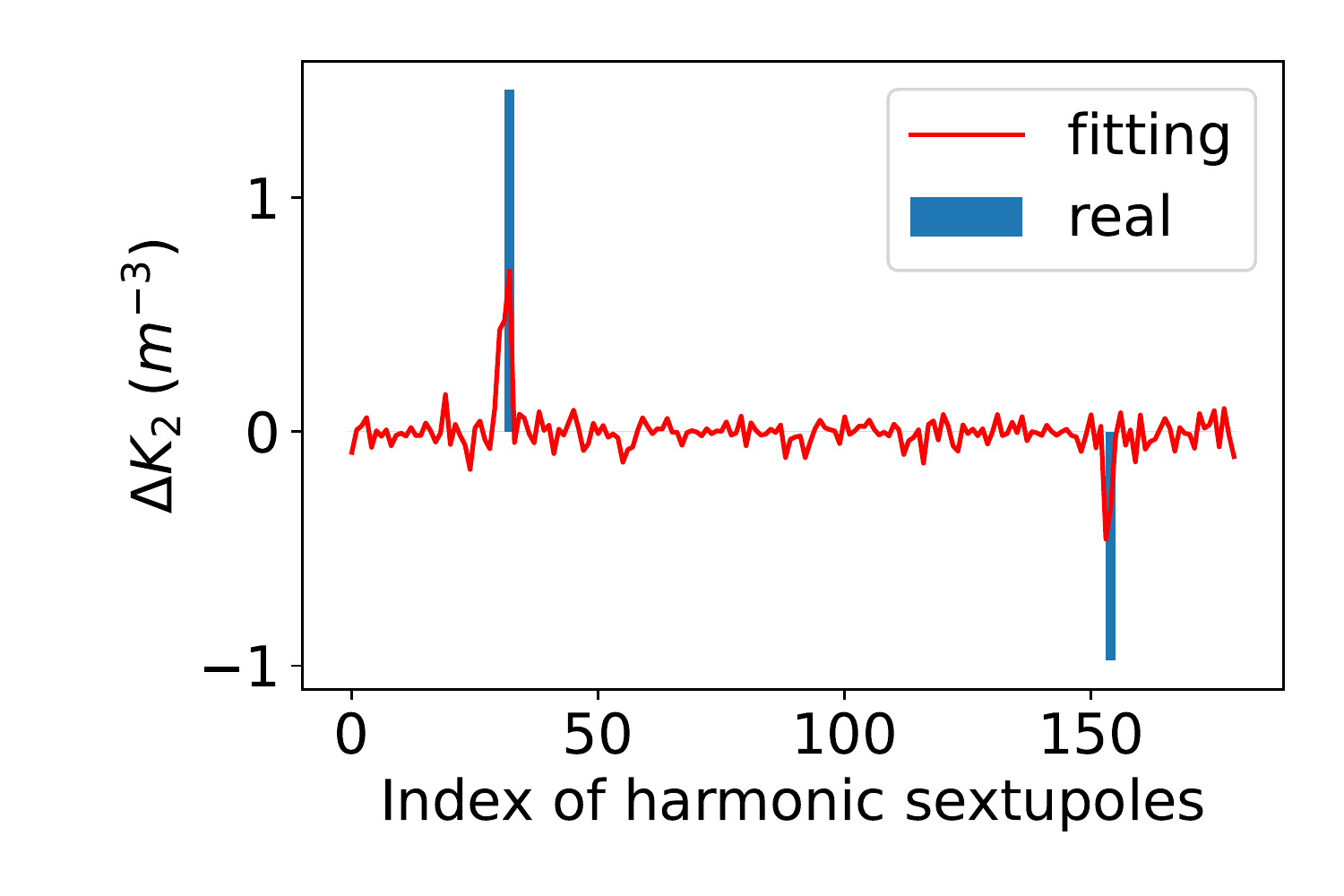}
      \caption{\label{fig:double_dK2} Correction scheme for two isolated
        sextupole errors obtained by the linear regression
        algorithm. Despite high degeneracy among sextupoles, two error
        sources are precisely localized.}
    \end{figure}

    The correction scheme obtained with Eq.~\eqref{eq:correction} for 180
    harmonic sextupoles is shown in Fig.~\ref{fig:double_dK2}. Due to the
    high degeneracy among the neighboring sextupoles, the scheme doesn't
    reproduce the original error distributions. Instead, they spread to
    their neighbors. Nevertheless, the errors were localized, and after
    applying the correction scheme, the nonlinear optics were recovered as
    illustrated in Fig.~\ref{fig:double_distort}.
  
  \subsection{Case 2: normally distributed errors}

\begin{figure}[!ht]
      \centering \includegraphics[width=\columnwidth]{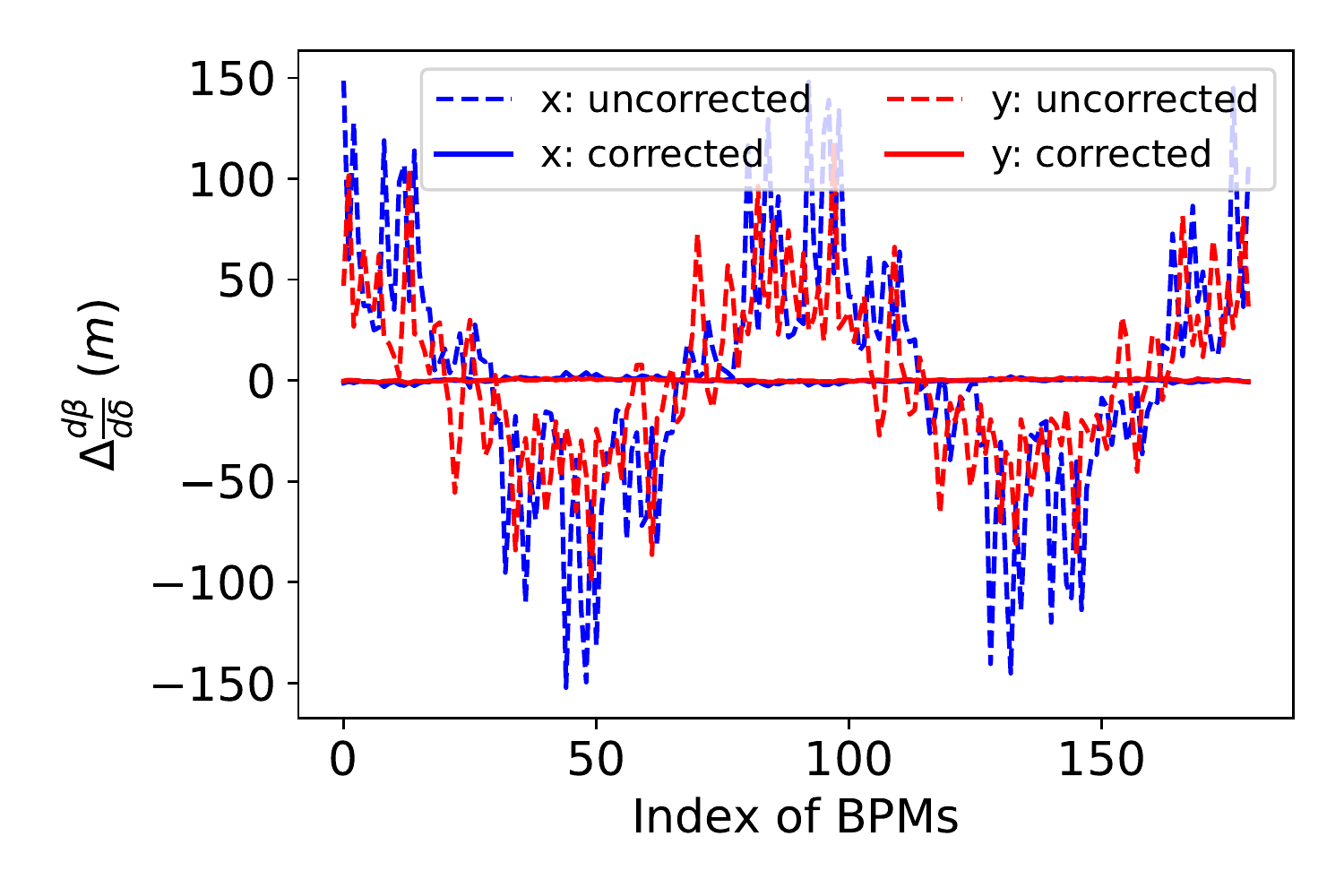}
      \caption{\label{fig:all_distort} Distortions of the nonlinear optics
        with 180 randomly added errors. The dashed lines are uncorrected
        distortions, and the solid lines represent the distortions after
        correction.}
    \end{figure}

\begin{figure}[!ht]
      \centering \includegraphics[width=\columnwidth]{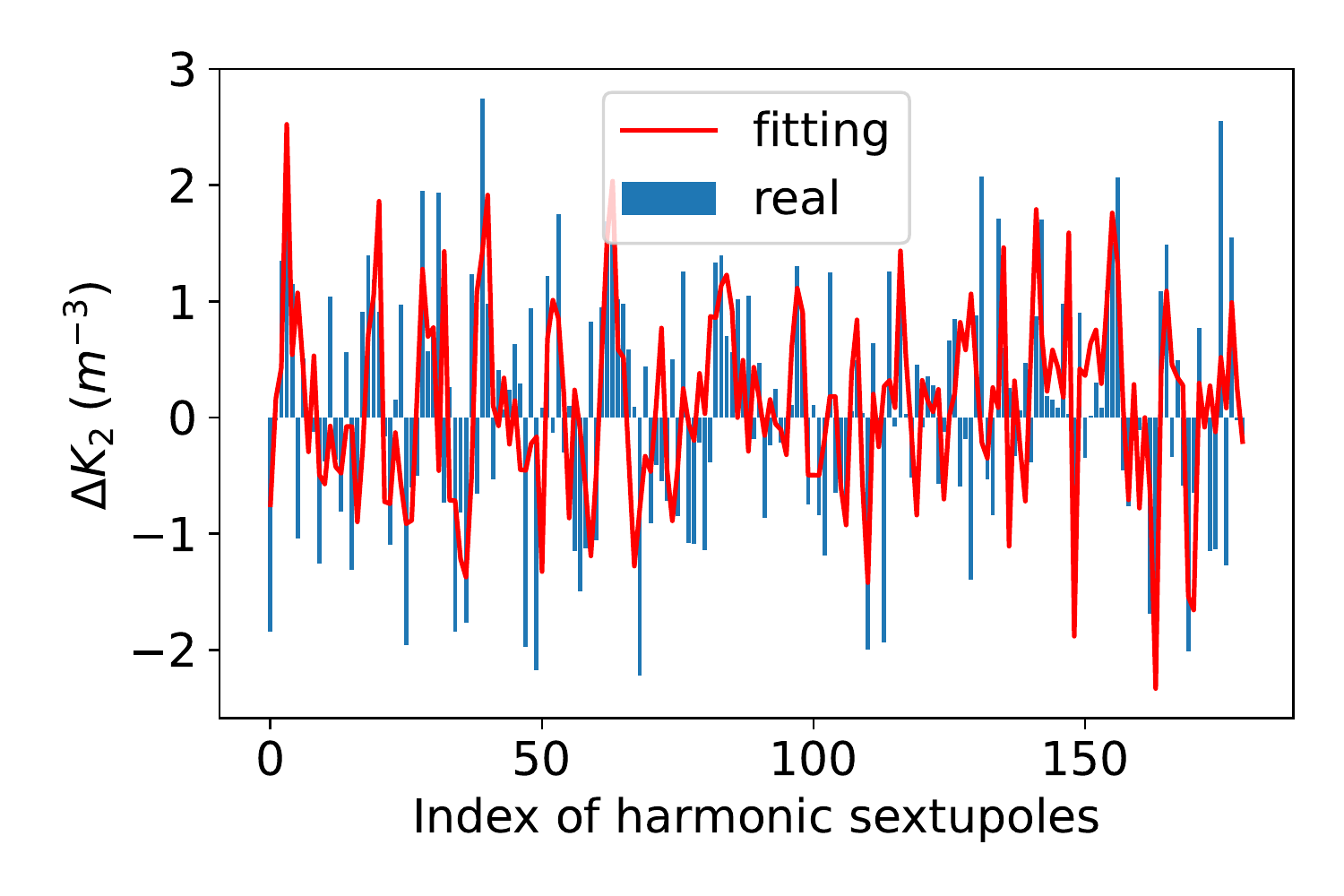}
      \caption{\label{fig:all_dK2} Comparison of the added errors (labeled
        as ``real'') on 180 harmonic sextupoles and the obtained
        correction scheme (shown with red lines labeled as ``fitting'').}
    \end{figure}

    In this case, random distributed errors
    on all 180 harmonic sextupoles are introduced and the distortion of
    off-energy optics are computed. Then the same correction
    procedure is employed. For comparison, the optics distortions before and after
    correction, and the real error distributions and computed correction
    scheme are illustrated in Fig.~\ref{fig:all_distort} and
    Fig.~\ref{fig:all_dK2}), respectively.

    In both cases, as seen in Fig.~\ref{fig:double_dK2} and
    \ref{fig:all_dK2}, the obtained correction schemes only approximately
    follow the real errors that were added in advance. This is due to the
    strong degeneracy that exists among sextupoles in the NSLS-II
    lattice. Minor imperfections of the BPMs, and other errors can also
    result in some degeneracy. However, the distortion of nonlinear optics
    can still be well corrected in Fig.~\ref{fig:double_distort} and
    \ref{fig:all_distort}. It is also worth mentioning that the dependence
    of nonlinear optics on sextupoles is not purely linear, therefore, an
    iterative correction might be necessary in online measurements.

  \subsection{Improvement on dynamic aperture degradation}

\begin{figure}[!ht]
      \centering \includegraphics[width=\columnwidth]{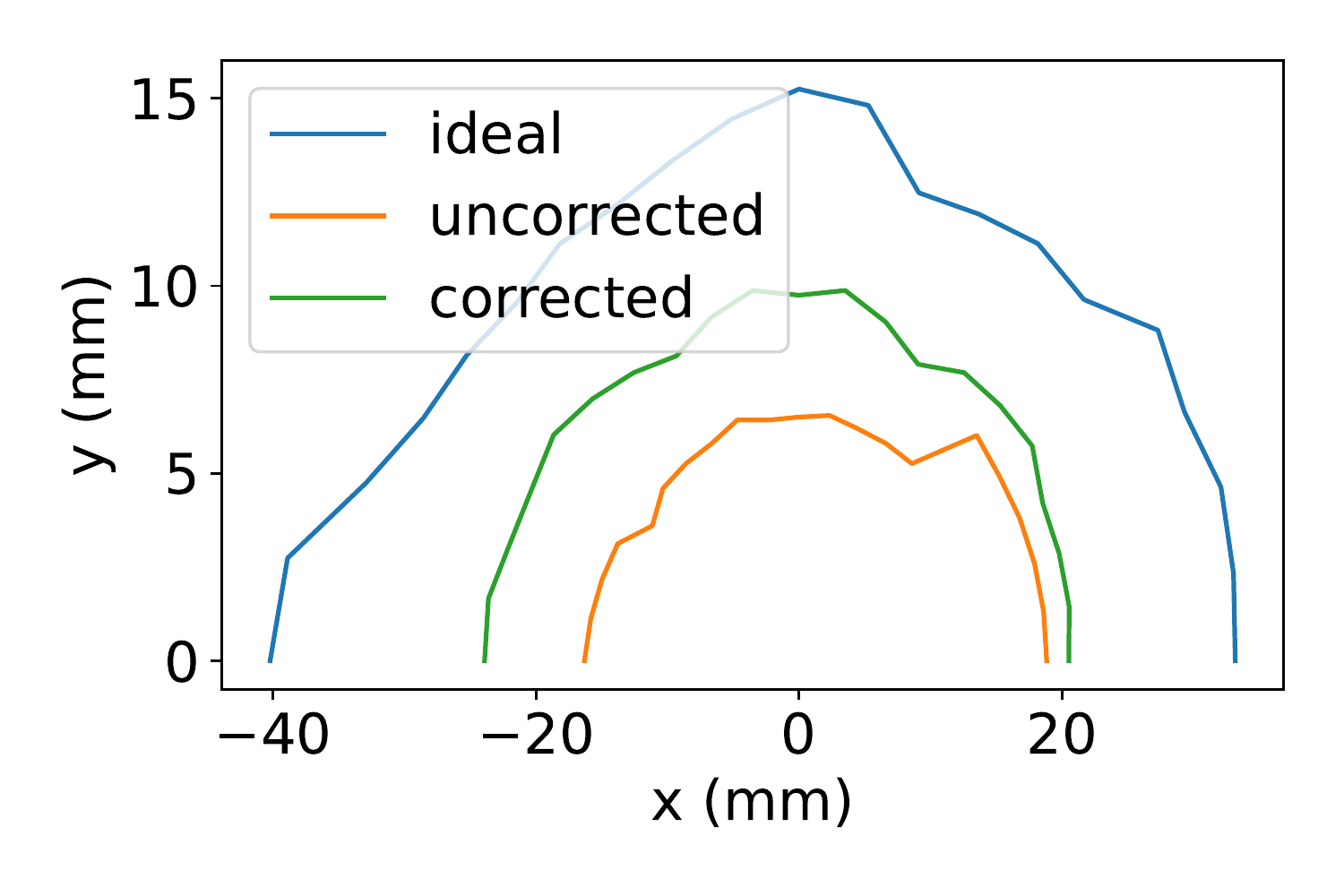}
      \caption{\label{fig:dyap} Comparison of the dynamic apertures for
        the ideal machine, and uncorrected/corrected nonlinear
        lattices. Although the degraded dynamic aperture (yellow line)
        could not be well recovered to that of the ideal machine (blue
        line), significant improvement (green line) was achieved, through
        correcting the distorted nonlinear optics.}
    \end{figure}

    As observed in the previous simulations, strong degeneracy among
    sextupoles prevents reproducing the real error distributions
    accurately. It is because, on the NSLS-II ring, every three harmonic
    sextupoles on the same girder are closely assembled. In our case, what
    is actually corrected is the distorted nonlinear optics dependence on
    beam energy deviations seen by the BPMs. The correction scheme based
    on the BPM observations, therefore, might only be able to recover the
    optics distortion rather than the dynamic aperture. To illustrate
    this, the dynamic apertures of the ideal machine, and
    uncorrected/corrected nonlinear lattices for the $2^{nd}$ simulation
    were computed for comparison (Fig.~\ref{fig:dyap}). Although the
    degraded dynamic aperture due to sextupole errors could not be fully
    recovered through the correction scheme, a significant improvement was
    achieved. Such improvement is the main purpose of calibrating and
    correcting the distorted nonlinear optics. If we could distinguish
    between the degeneracy among the sextupoles, further improvement could
    be made. This topic is slightly beyond the scope of this paper,
    however, but worth more study.

\section{\label{sect:meas}Measurements}

  \subsection{Two-stage measurements}
\begin{figure}[!ht]
      \centering
      \includegraphics[width=\columnwidth]{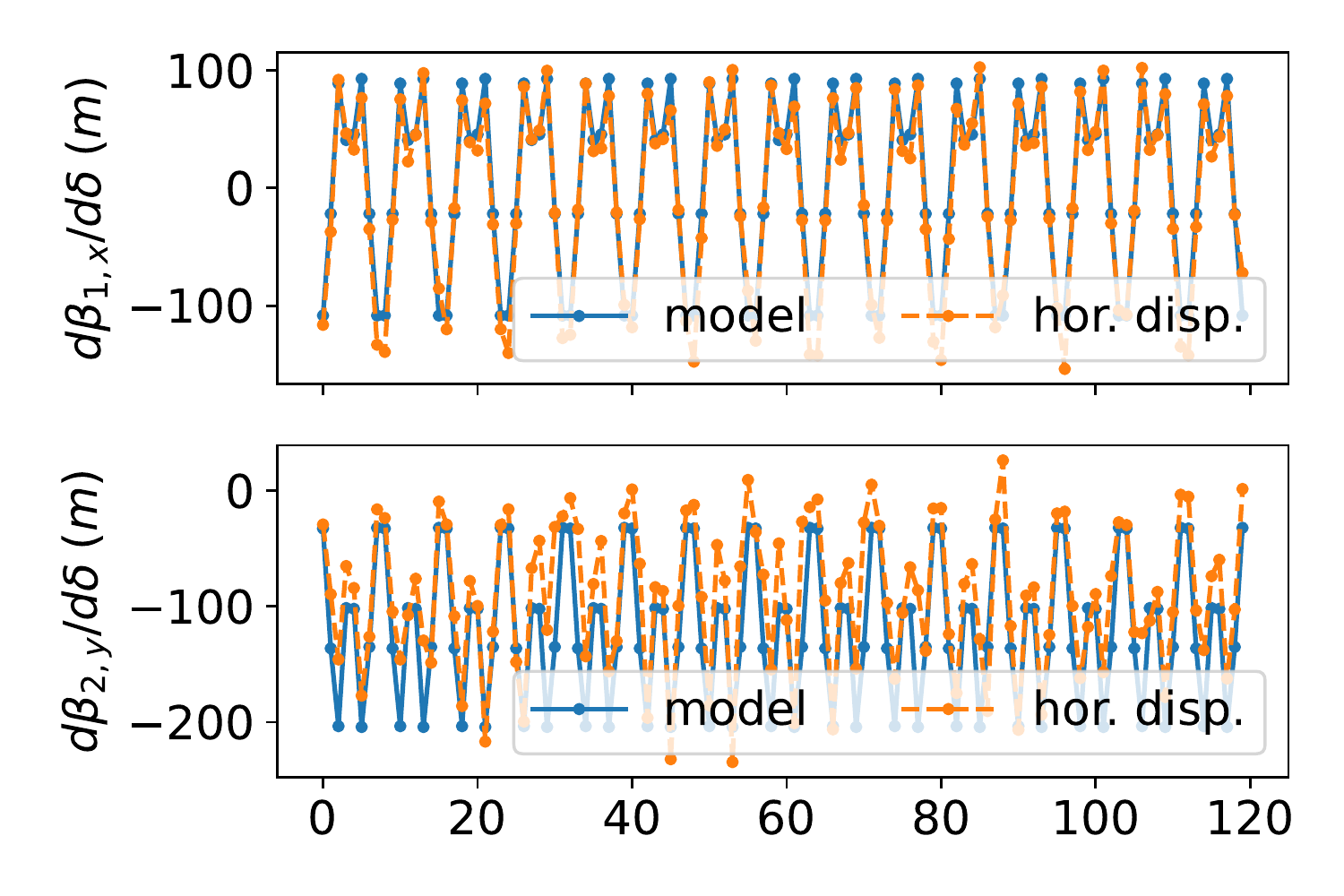}
      \includegraphics[width=\columnwidth]{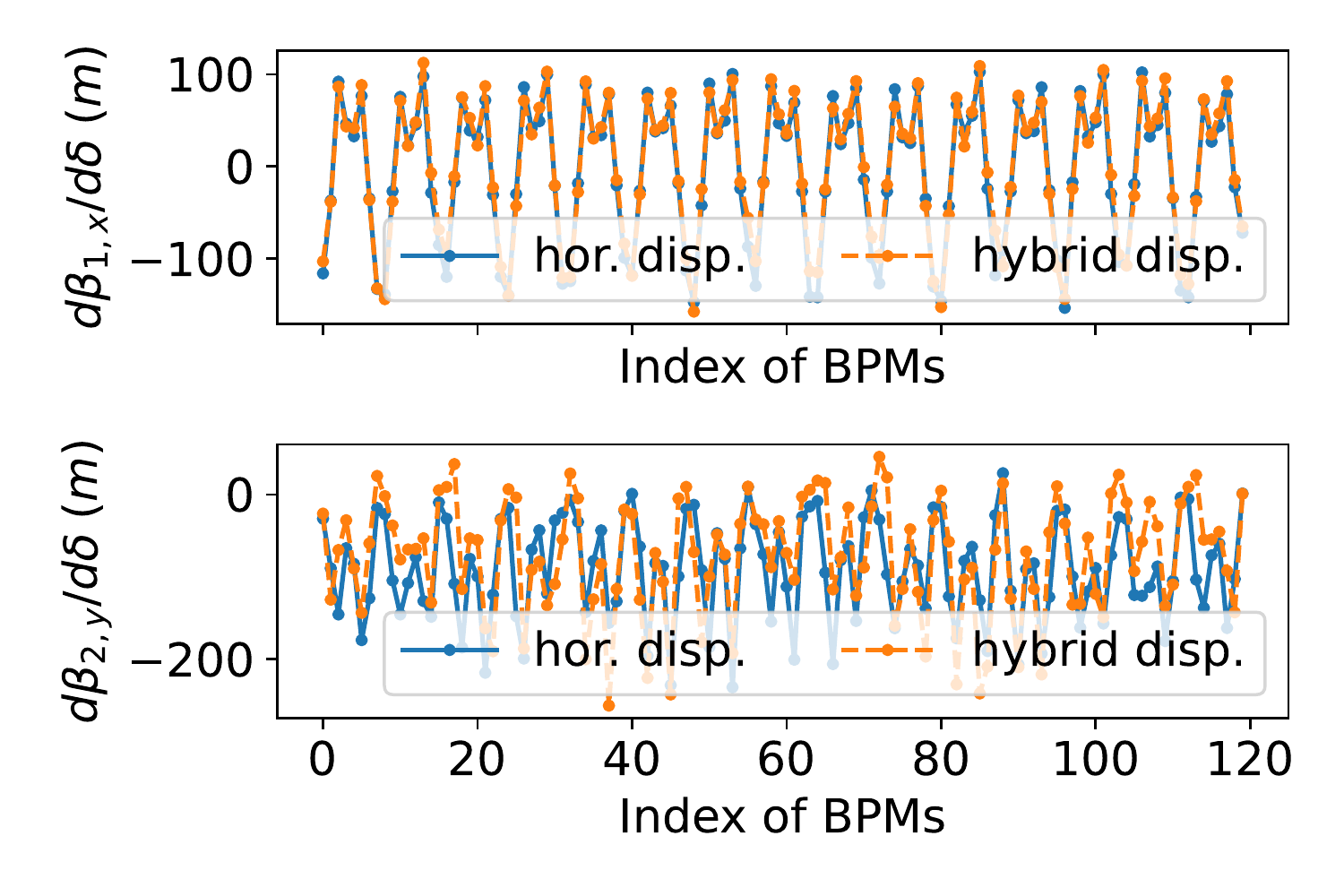}
      \caption{\label{fig:two_stage_meas} Measured off-energy optics at
        two stages. Top: off-energy optics based on the ideal model and
        online measurements from horizontal dispersive orbits which can be
        used to calibrate chromatic sextupoles. Bottom: online
        measurements of off-energy optics from horizontal and hybrid
        dispersive orbits, which can be used to calibrate harmonic
        sextupoles.}
    \end{figure}

    A two-stage proof-of-principle through online calibration of sextupole
    errors was implemented at the NSLS-II storage ring. As the sextupoles
    are powered in series, the sextupoles lack individual configurability.
    Therefore, no actual nonlinear optics correction can be implemented
    with these limitations. For stage-1, we calibrated 90 chromatic
    sextupoles with existing techniques. First, spurious vertical
    dispersion was minimized using 15 chromatic skew quadrupoles, and the
    global linear coupling was well corrected with another 15
    non-dispersive skew quadrupoles. The $\frac{d\beta}{d\delta}$ seen by
    the BPMs were measured from horizontal dispersive orbits through
    varying the beam energies. By comparing the measured nonlinear optics
    against the design model, the chromatic sextupole errors (red bars in
    Fig.~\ref{fig:two_stage_meas}) were obtained using the model response
    matrices, and then incorporated into the lattice model. The updated
    model would be used as the reference for the stage-2 calibration.

\begin{figure}[!ht]
    \centering \includegraphics[width=\columnwidth]{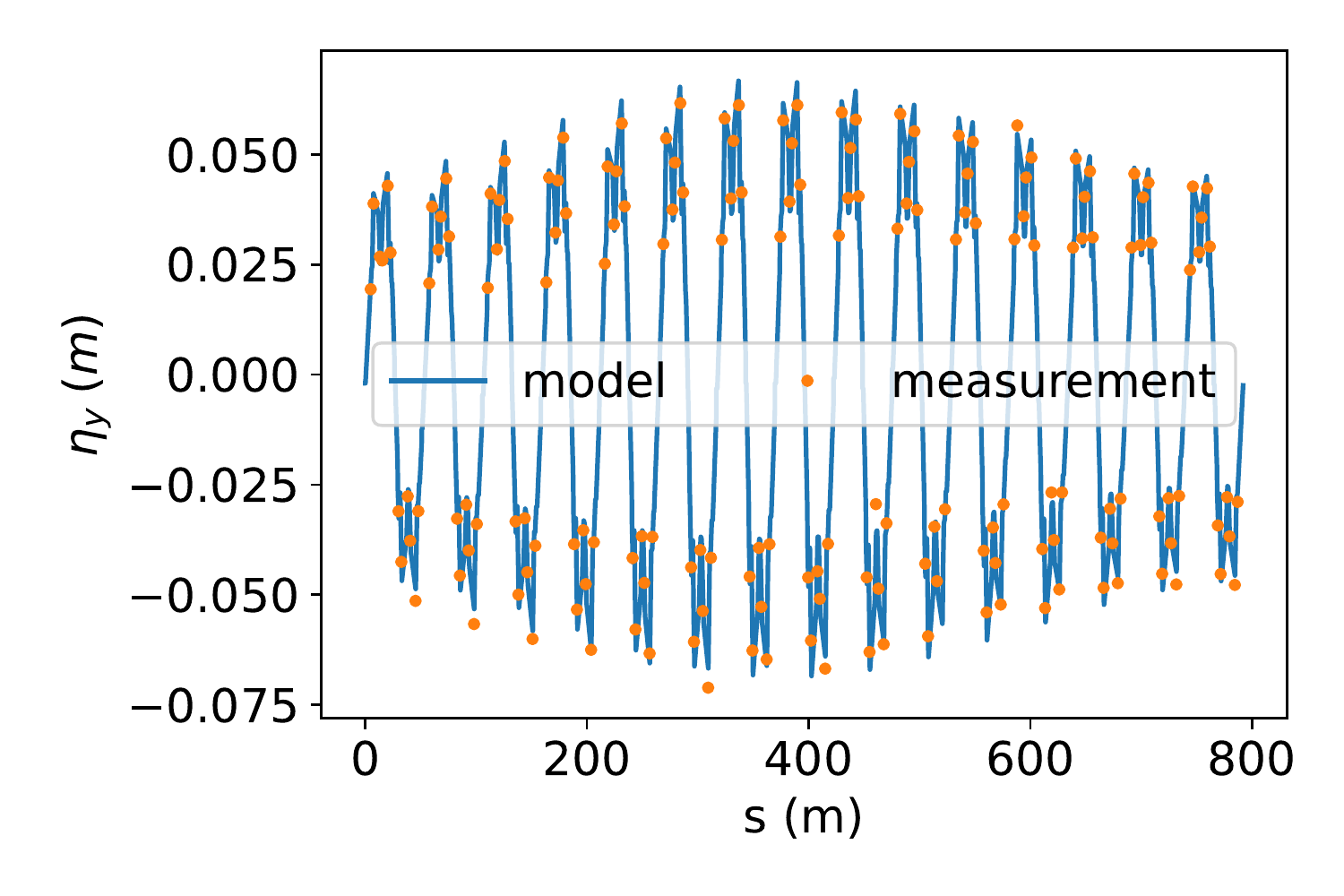}
    \caption{\label{fig:vdisp} Measured vertical dispersion seen by the BPMs,
      and its reproduction with the lattice model.}
    \end{figure}

    For stage-2, a vertical dispersion wave was generated with 15
    dispersive skew quadrupoles to their maximum capacity. Based on the
    measured dispersion, the 15 skew quadrupole settings and the vertical
    dispersion at the BPMs were reproduced with the lattice model as
    illustrated in Fig.~\ref{fig:vdisp}. To achieve greater accuracy, a
    large amplitude vertical dispersion wave is preferred. However, it is
    limited by the capacity of the skew quadrupole power supply. Under the
    current configuration, $\sim0.05\;m$ is the maximum amplitude that can
    be generated. The $\frac{d\beta}{d\delta}$ seen by the BPMs were
    re-measured, but from hybrid dispersive orbits this time. With the
    updated lattice model (incorporated with skew quadrupoles and
    chromatic sextupole errors) as the new reference, 180 harmonic
    sextupole errors were obtained (blue bars in
    Fig.~\ref{fig:two_stage_meas}.

\begin{figure}[!ht]
    \centering \includegraphics[width=\columnwidth]{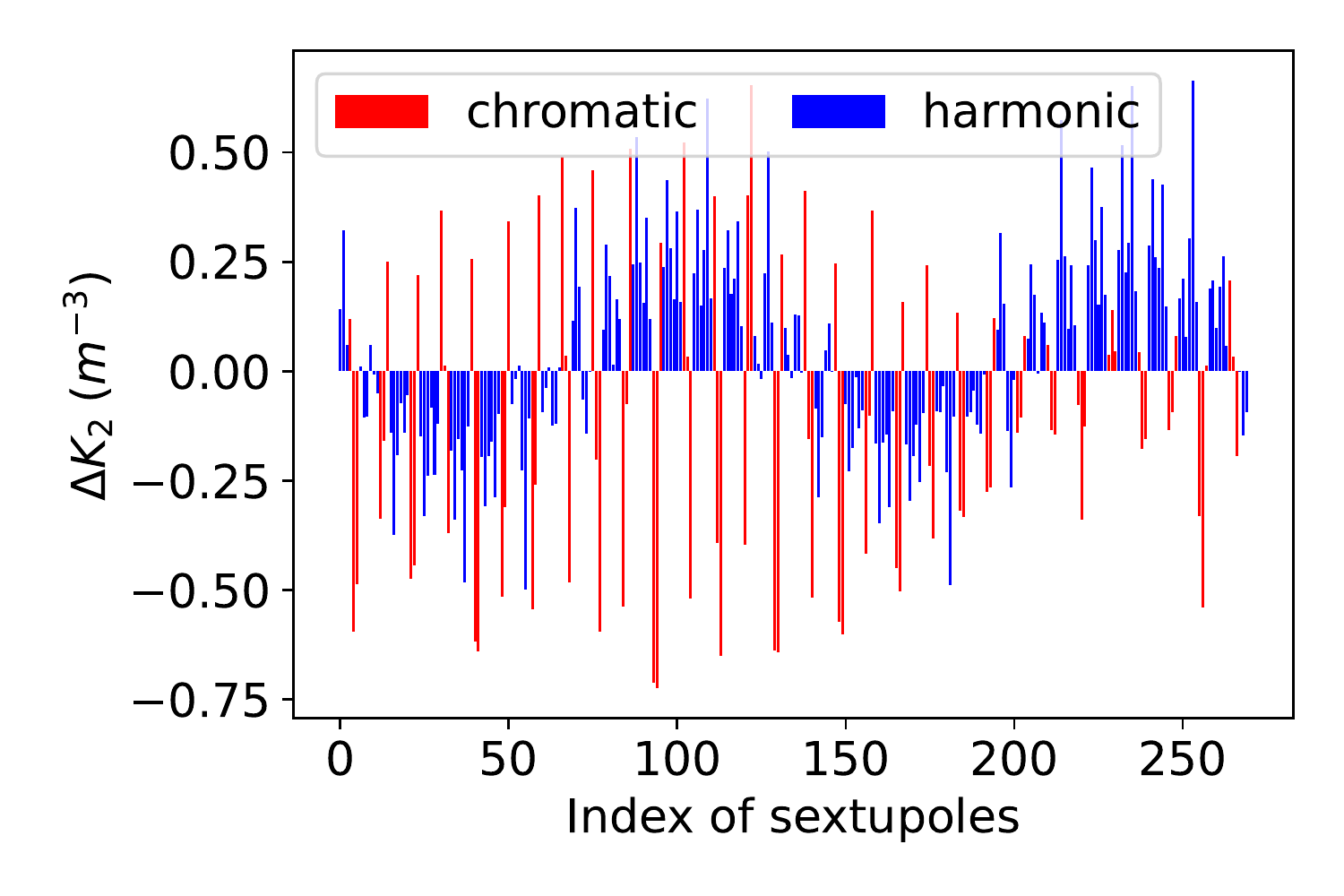}
    \caption{\label{fig:measdK2} Calibrated sextupole errors from the
      two-stage measurements. The chromatic sextupoles marked with red
      bars are from the first stage, while the harmonic ones (blue bars)
      are from the second stage.}
    \end{figure}

    The off-energy optics ($\beta$-functions) were measured with different
    RF frequencies $f_0+\Delta f$, i.e., different energy
    $\delta=-\frac{1}{\alpha_c}\frac{\Delta f}{f_0}$, with $\alpha_c$ the
    momentum compaction factor, and $f_0$ the nominal RF oscillator
    frequency for on-momentum electrons. Using Eq.~\eqref{eq:correction},
    the sextupole errors were calibrated as illustrated in
    Fig.~\ref{fig:measdK2}.

  \subsection{Validation of stage-2 measurement}

\begin{figure}[!ht]
    \centering \includegraphics[width=\columnwidth]{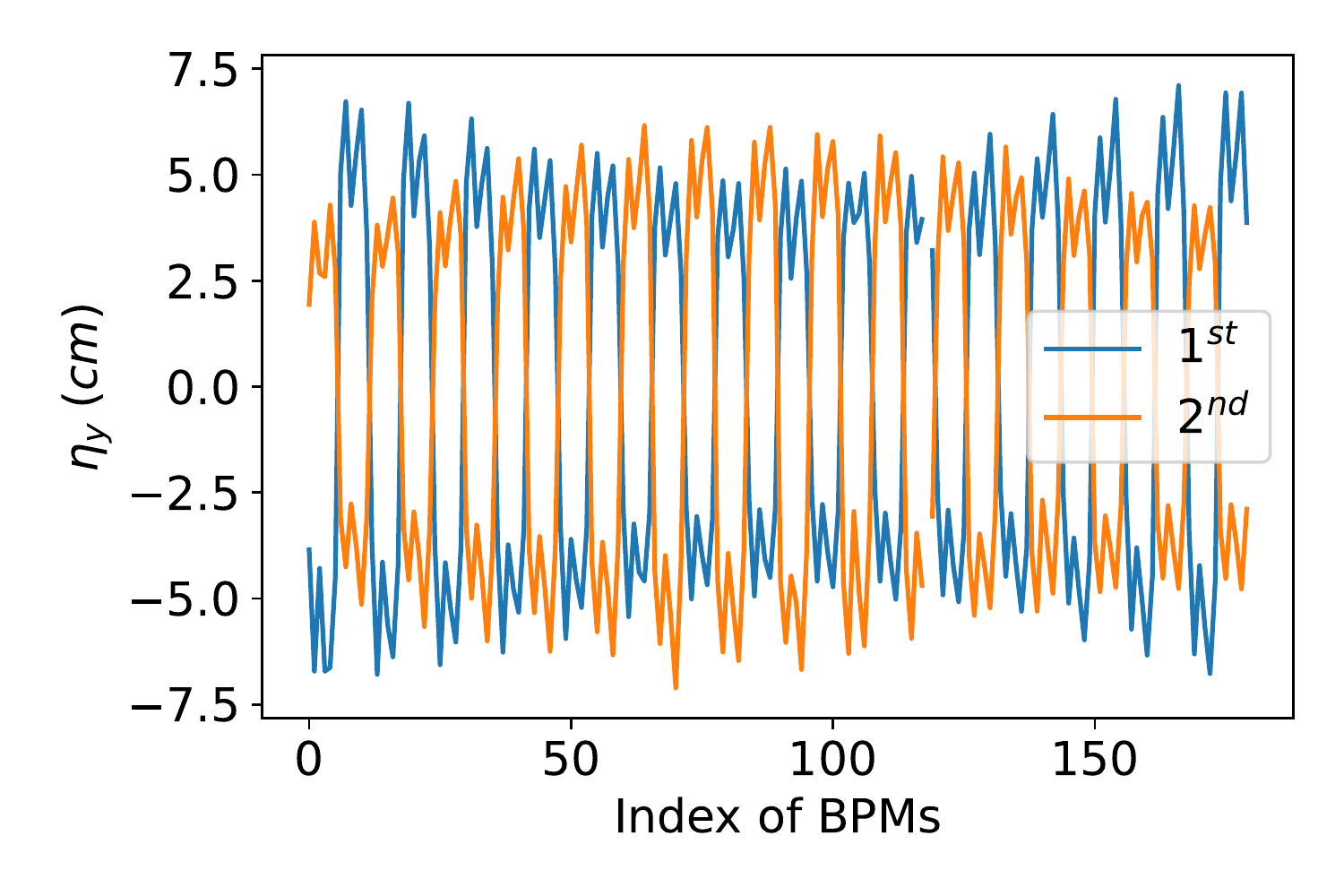}
    \caption{\label{fig:flipDisp} Out-of-phase vertical dispersion waves
      were generated with flipped skew quadrupole polarities.}
    \end{figure}

\begin{figure}[!ht]
    \centering \includegraphics[width=\columnwidth]{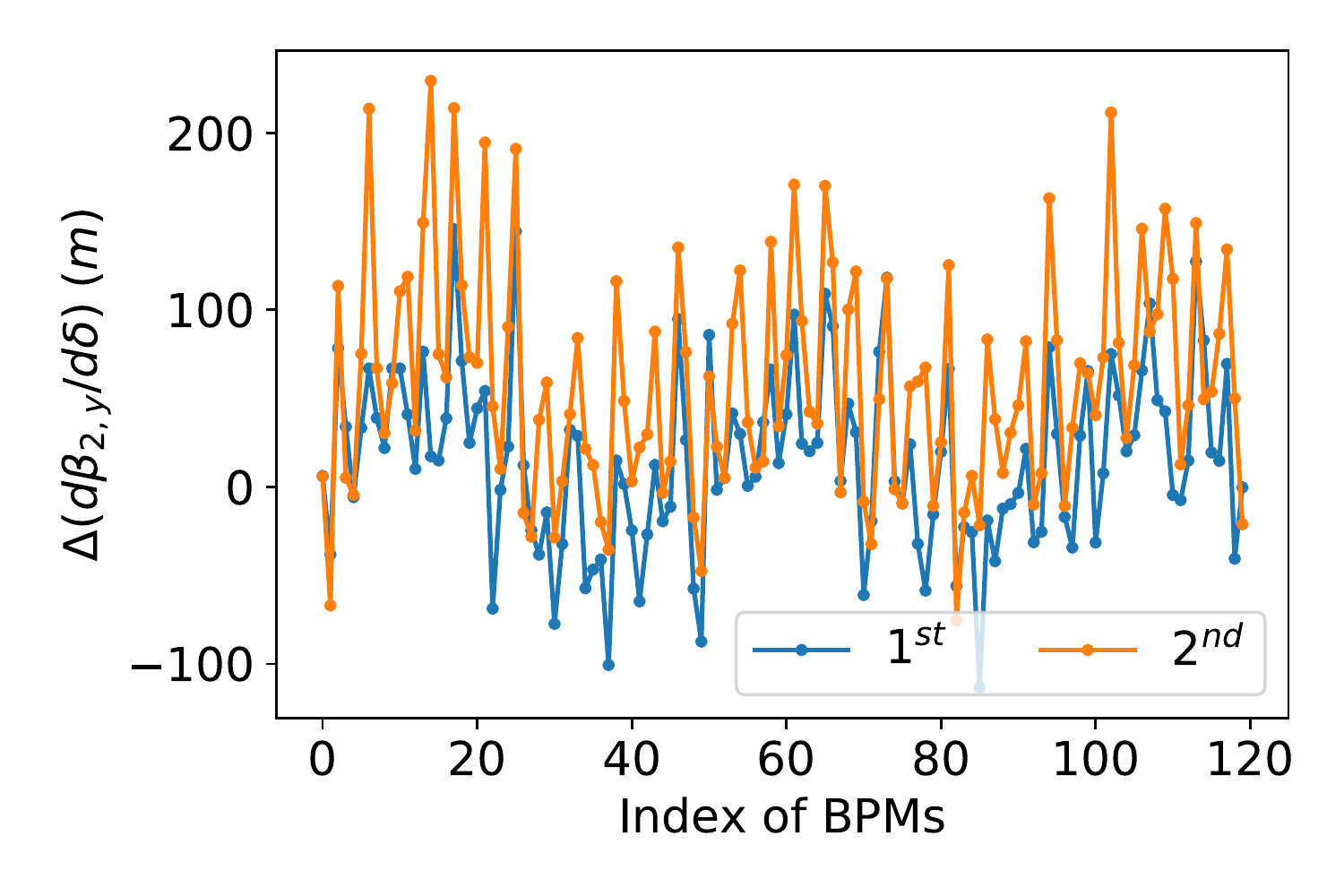}
    \caption{\label{fig:flipErr} Similar off-energy optics distortions
      were observed when the hybrid dispersive orbit was flipped in the
      vertical plane.}
    \end{figure}

    Skew quadrupoles used as correctors are usually operated with dual
    polarity power supplies. By flipping skew quadrupole polarities,
    hybrid dispersive orbits are also flipped in the vertical plane. In
    Appendix~\ref{sect:flip}, we prove that the off-energy optics
    dependence on sextupoles in the flipped orbit remains unchanged when
    the Ripken parameterization is used. Therefore, we can repeat the
    stage-2 measurement on the flipped vertical dispersive orbits as a
    validation. In Fig.~\ref{fig:flipDisp}, two out-of-phase dispersion
    waves were obtained by flipping all skew quadrupole outputs $K_1$ from
    $-0.035\;m^{-2}$ to $0.035\;m^{-2}$. On the flipped vertical
    dispersive orbits, stage-2 measurements of off-energy optics changed
    with respect to stage-1, were repeated as illustrated in
    Fig.~\ref{fig:flipErr}. A similar pattern can be recognized in two
    independently measured optics distortions. This indicates that the
    dependence of off-energy optics on harmonic sextupoles is measurable
    on the hybrid dispersive orbit, although the precision is limited by
    the low capacity of skew quadrupole power supplies.
    
\section{\label{sect:hardware}Requirements on power supplies of magnets}

  In this section, we estimate the requirements of the power supplies for
  the skew quadrupoles and sextupoles to make this calibration and
  correction practical, specifically on the NSLS-II ring. First, sextupoles
  need to be powered independently, or equipped with back-leg windings to
  allow individual corrections. Second, the skew quadrupoles that generate
  the vertical dispersion wave should be sufficiently strong for better
  resolution. Next, we use the NSLS-II lattice to estimate the required
  skew quadrupole strengths.

  The measurement accuracy of $\beta$-functions using turn-by-turn data
  was found at the level of about $0.03\;m$ on the NSLS-II storage
  ring. If we adjust the RF frequency by $\pm1,000\;Hz$, the corresponding
  energy deviation is $\Delta\delta=-\frac{1}{\alpha_c}\frac{\Delta
    f}{f_0}\approx1.0\%$ with $\alpha_c$ momentum compaction factor. To
  resolve a single harmonic sextupole error to the level of 1 unit $\Delta
  K_2=1\;m^{-3}$, the magnitude of $\frac{d}{dK_2}\frac{d\beta}{d\delta}$
  needs to be greater than $3\;m^4$. To generate such a strong dependence,
  the amplitude of the vertical dispersion wave is required to be greater
  than $0.1\;m$. In order to couple $0.1\;m$ dispersion to the vertical
  plane, $K_{1,sm}\ge0.07\;m^{-2}$ must be applied to all 15 skew
  quadrupoles. Therefore, strong skew quadrupoles were chosen to identify
  two isolated errors in our simulations. Although a $\frac{\partial
    B_x}{\partial x}=0.70\;T\cdot m^{-1}$ quadrupolar gradient is still
  quite weak, it already exceeds the capacity of our skew quadrupole power
  supplies $g_{2,max}=0.35\;T\cdot m^{-1}$. In other words, stronger skew
  quadrupoles are needed to generate larger vertical dispersion for better
  resolution. As all 270 sextupoles have some associated errors, the
  accumulated magnitude of $\Delta\frac{d\beta}{d\delta}$ are at the level
  of 100-200 meters, which allows us to calibrate the approximate error
  distribution in Sect.~\ref{sect:meas}.

  The above estimation doesn't consider even higher orders of nonlinear
  optics from off-energy orbits. Once the higher order nonlinearities
  $\frac{d^n\beta}{d\delta^n}$ with $n\ge2$ appear, we couldn't improve
  the measurement accuracy by increasing beam energy off-sets. On the
  other hand, increasing skew quadrupole strengths to generate higher
  vertical dispersive orbits can significantly improve the sensitivity of
  off-energy optics to sextupole settings without introducing too much
  nonlinearity. Therefore, having sufficiently strong skew quadrupoles
  should be considered a necessary condition for this technique.
  
\section{\label{sect:summary}Summary}

  We expanded the capability of the technique for measuring nonlinear
  optics distortions from off-energy orbits to account for the harmonic
  sextupole contribution. Using hybrid dispersive off-energy nonlinear
  optics, the errors of the harmonic sextupoles can be measured. The
  corresponding correction can be more effectively implemented if they are
  independently configurable. A practical benefit of our expanded method
  is that a considerable amount of vertical dispersion can be generated
  with weak skew quadrupoles. Meanwhile, because the original lattice is
  already weakly coupled, its optics properties can still be well
  maintained. Thus far, only sextupole calibration was considered in our
  studies, and higher order nonlinear magnets, such as octupoles, we have
  not yet investigated. This technique might be applicable if their
  contributions were sufficiently strong.
  
\begin{acknowledgments}
  We would like to thank the collaborative and productive discussion with
  Dr. D. Olsson (MAX IV, Lund Uni.), Prof. Y. Hao (MSU), and some NSLS-II
  colleagues, Dr. J. Choi, Dr. Y. Hidaka, Dr. M. Song, Dr. G. Tiwari,
  Dr. X. Yang, et al. This research used resources of the National
  Synchrotron Light Source II, a U.S. Department of Energy (DOE) Office of
  Science User Facility operated for the DOE Office of Science by
  Brookhaven National Laboratory under Contract No. DE-SC0012704.
\end{acknowledgments}

\section*{\label{sect:flip}Appendix: flipped dispersive orbit
  in the vertical plane}

  The hybrid dispersive orbit can be flipped only in the vertical plane by
  changing the skew quadrupole polarities. Usually skew quadrupoles used
  for the correction purposes are operated with dual polarity power
  supplies. In this appendix, we prove that an exact flipping of the
  vertical dispersive orbit doesn't change the dependence of off-energy
  optics on sextupoles which can be used to validate online measurements.

\subsection{Transfer matrix with single skew quadrupole}

  The $4\times4$ transfer matrix of a normal quadrupole reads as
  \begin{equation}
    T_n=\left[
      \begin{matrix}
	T_x & \mathbf{0}\\
	\mathbf{0} & T_y
      \end{matrix}
      \right]
  \end{equation}
  with two $2\times2$ zero blocks $\mathbf{0}$ as its off-diagonal
  matrices.

  After rotating it by $\frac{\pi}{4}$ around the longitudinal axis, it
  becomes a skew with a transfer matrix
  \begin{equation}
    T_s=R_z\left( \frac{\pi}{4}\right)T_nR_z\left(-\frac{\pi}{4}\right)
    =\left[
      \renewcommand{\arraystretch}{1.5}
      \begin{matrix}
	\frac{T_x+T_y}{2} & \frac{-T_x+T_y}{2}\\
	\frac{-T_x+T_y}{2} & \frac{T_x+T_y}{2}
      \end{matrix}
      \right]
    \label{eq:Ts}
  \end{equation}
  where $R_z\left( \theta \right)$ represents the transfer matrix of a pure
  rotation with an angle of $\theta$.
  
  Assuming there is only one skew quadrupole inside a periodic lattice
  cell with the layout ``normal section 1 -- skew quad -- normal section
  2'', the transfer matrix of the whole section is
  \begin{equation}
    M_t=
    \left[
      \begin{matrix}
	M_{x,2} & \mathbf{0}\\
	\mathbf{0} & M_{y,2}
      \end{matrix}
      \right]
    T_s
    \left[
      \begin{matrix}
	M_{x,1} & \mathbf{0}\\
	\mathbf{0} & M_{y,1}
      \end{matrix}
      \right]
    =\left[
      \begin{matrix}
	A & B\\
	C & D
      \end{matrix}
      \right]
    \label{eq:Mt}
  \end{equation}

  When the skew quadrupole polarity is flipped from $K_{1,s}$ to
  $-K_{1,s}$, the new transfer matrix can be obtained by swapping $T_x$
  and $T_y$ in Eq.~(\ref{eq:Ts}),
  \begin{equation}
    T_s'=\left[
      \renewcommand{\arraystretch}{1.5}
      \begin{matrix}
	\frac{T_y+T_x}{2} & \frac{-T_y+T_x}{2}\\
	\frac{-T_y+T_x}{2} & \frac{T_y+T_x}{2}
      \end{matrix}
      \right]=
    ST_sS^{-1}
    \label{eq:Ts1}
  \end{equation}
  with
  \begin{equation}
    S=
    \left[
      \begin{matrix}
	\mathbf{1} & \mathbf{0}\\
	\mathbf{0} & -\mathbf{1}
      \end{matrix}
      \right]
  \end{equation}
  where $\mathbf{1}$ is the $2\times2$ identity matrix. The transfer
  matrix of the whole section then becomes
  \begin{widetext}
  \begin{equation}
    \begin{aligned}
      M_t'&=
      \left[
	\begin{matrix}
	  M_{x,2} & \mathbf{0}\\
	  \mathbf{0} & M_{y,2}
	\end{matrix}
        \right]
      T_s'
      \left[
	\begin{matrix}
	  M_{x,1} & \mathbf{0}\\
	  \mathbf{0} & M_{y,1}
	\end{matrix}
        \right]
	=S\left( 
	S^{-1}
	\left[ 
		\begin{matrix}
			M_{x,2} & \mathbf{0}\\
			\mathbf{0} & M_{y,2}
		\end{matrix}
	\right]S
	\right)
	T_s
	\left(  
	S^{-1}
	\left[ 
		\begin{matrix}
			M_{x,1} & \mathbf{0}\\
			\mathbf{0} & M_{y,1}
		\end{matrix}
	\right]S
	\right)
	S^{-1}\\
      &=
      S\left[
	\begin{matrix}
	  A & B\\
	  C & D
	\end{matrix}
        \right]S^{-1}
      =\left[
	\begin{matrix}
	  A & -B\\
	  -C & D
	\end{matrix}
        \right]
    \end{aligned}
    \label{eq:Mt1}
  \end{equation}
  \end{widetext}

  By comparing Eq.~(\ref{eq:Mt}) and Eq.~(\ref{eq:Mt1}), we can conclude
  that when a periodic cell contains only one skew quadrupole, only the
  signs of the off-diagonal blocks $B$ and $C$ flip when the skew
  quadrupole polarity is flipped.

  \subsection{Ripken Twiss functions with single flipped skew}
  Using the Ripken parameterization\cite{willeke1989,lebedev2010} on
  $M_t$, the transfer matrix is diagonalized respectively as
  \begin{equation}
      M_t=UVU^{-1},
      \label{eq:MtRipken}
  \end{equation}
  with
  \begin{widetext}
  \begin{equation}
      V=\left[
	\begin{matrix}
	  \cos\mu_1 & \sin\mu_1 & 0 & 0\\
	  -\sin\mu_1 & \cos\mu_1 & 0 & 0\\
	  0 & 0 & \cos\mu_2 & \sin\mu_2\\
	  0 & 0 & -\sin\mu_2 & \cos\mu_2
	\end{matrix}
        \right],
      U=\left[
	\begin{matrix}
	  \sqrt{\beta_{1x}} & 0 & \sqrt{\beta_{2x}}\cos v_2 &
          -\sqrt{\beta_{2x}}\sin
          v_2\\ -\frac{\alpha_{1x}}{\sqrt{\beta_{1x}}} &
          \frac{1-u}{\beta_{1x}} & \frac{-\alpha_{2x}\cos v_2+u\sin
            v_2}{\sqrt{\beta_{2x}}} & \frac{\alpha_{2x}\sin v_2+u\cos
            v_2}{\sqrt{\beta_{2x}}}\\ \sqrt{\beta_{1y}}\cos v_1 &
          -\sqrt{\beta_{1y}}\sin v_1 & \sqrt{\beta_{2y}} &
          0\\ \frac{-\alpha_{1y}\cos v_1 + u\sin v_1}{\sqrt{\beta_{1y}}} &
          \frac{\alpha_{1y}\sin v_1 +u\cos v_1}{\sqrt{\beta_{1y}}} &
          -\frac{\alpha_{2y}}{\sqrt{\beta_{2y}}} &
          \frac{1-u}{\sqrt{\beta_{2y}}}
	\end{matrix}
        \right]
  \end{equation}
  \end{widetext}
  where $\beta,\alpha,\mu,\nu,u$ are defined in \cite{lebedev2010}.

  Now we implement the same parameterization on $M_t'$ with a flipped skew
  quadrupole polarity,
  \begin{equation}
    \begin{aligned}
      M_t'&=SM_tS^{-1}=SUVU^{-1}S^{-1}\\
      &=\left( SUS^{-1} \right)\left( SVS^{-1}
      \right)\left( SUS^{-1} \right)^{-1}\\
      &=U'V'U'^{-1}
      \label{eq:Mt1Ripken}
    \end{aligned}
  \end{equation}
  with
  \begin{equation}
    V'=V
    \label{eq:V1}
  \end{equation}
  and
  \begin{widetext}
    \begin{equation}
      U'=SUS^{-1}=\left[
        \begin{matrix}
          \sqrt{\beta_{1x}} & 0 & -\sqrt{\beta_{2x}}\cos v_2 &
          \sqrt{\beta_{2x}}\sin
          v_2\\ -\frac{\alpha_{1x}}{\sqrt{\beta_{1x}}} &
          \frac{1-u}{\beta_{1x}} & \frac{\alpha_{2x}\cos v_2-u\sin
            v_2}{\sqrt{\beta_{2x}}} & -\frac{\alpha_{2x}\sin v_2+u\cos
            v_2}{\sqrt{\beta_{2x}}}\\ -\sqrt{\beta_{1y}}\cos v_1 &
          \sqrt{\beta_{1y}}\sin v_1 & \sqrt{\beta_{2y}} &
          0\\ \frac{\alpha_{1y}\cos v_1 - u\sin v_1}{\sqrt{\beta_{1y}}} &
          -\frac{\alpha_{1y}\sin v_1 +u\cos v_1}{\sqrt{\beta_{1y}}} &
          -\frac{\alpha_{2y}}{\sqrt{\beta_{2y}}} &
          \frac{1-u}{\sqrt{\beta_{2y}}}
        \end{matrix}
	\right]
      \label{eq:U1}
    \end{equation}
  By comparing Eq.~(\ref{eq:MtRipken}) and Eq.~(\ref{eq:U1}), we get
  \begin{equation}
    \begin{gathered}
	\mu_1'=\mu_1,\qquad \mu_2'=\mu_2, \qquad u'=u,\qquad
        v_1'=v_1+\pi,\qquad v_2'=v_2+\pi\\ \beta_{1x}'=\beta_{1x},\qquad
        \beta_{1y}'=\beta_{1y},\qquad \beta_{2x}'=\beta_{2x},\qquad
        \beta_{2y}'=\beta_{2y},\\ \alpha_{1x}'=\alpha_{1x},\qquad
        \alpha_{1y}'=\alpha_{1y},\qquad \alpha_{2x}'=\alpha_{2x},\qquad
        \alpha_{2y}'=\alpha_{2y}
    \end{gathered}
  \end{equation}
  \end{widetext}
  Therefore, flipping the skew quadrupole polarity doesn't change the signs
  or values of $\beta$ and $\alpha$ functions in the Ripken
  parameterization.

\subsection{Off-energy optics dependence on sextupoles with flipped dispersive orbit}
  Now we consider the off-energy optics dependence on sextupoles with the
  flipped dispersive orbit. When an off-energy particle passes through a
  sextupole with a vertical offset $y_0$, it sees a skew quadrupolar
  component, which is proportional to $K_2Ly_0 = K_2L\eta_y\delta$ with
  $K_2L$ as the sextupole’s effective field integral, and $\delta$ as the
  particle momentum offset. On the flipped vertical dispersion orbit, the
  one-turn transfer matrix, which is composed of a sequence of matrices
  (each of them has only one sextupole included)
  \begin{equation}
    M =
    \left[
    \begin{matrix}
      A & B\\
      C & D
    \end{matrix}
    \right]
    =
    \left[
    \begin{matrix}
      A_1 & B_1\\
      C_1 & D_1
    \end{matrix}
    \right]
    \left[
    \begin{matrix}
      A_2 & B_2\\
      C_2 & D_2
    \end{matrix}
    \right]
    \cdots
    \left[
    \begin{matrix}
      A_n & B_n\\
      C_n & D_n
    \end{matrix}
  \right]
  \end{equation}
  Here each section's transfer matrix $\left[\begin{matrix} A_i &
      B_i\\ C_i & D_i\end{matrix}\right]$ includes only one skew
  quadrupole or sextupole. After flipping vertical dispersion with skew
  quadrupoles, the one-turn matrix on the closed dispersive orbit with
  $-y_0$ can be obtained by following the same rule as in the presence of
  one skew quadrupole,
  \begin{equation}
  \begin{aligned}
    M' &=
    S
  \left[
    \begin{matrix}
      A_1 & B_1\\
      C_1 & D_1
    \end{matrix}
    \right]
  S^{-1}S
  \left[
    \begin{matrix}
      A_2 & B_2\\
      C_2 & D_2
    \end{matrix}
  \right]S^{-1}
  \cdots
  S
  \left[
    \begin{matrix}
      A_n & B_n\\
      C_n & D_n
    \end{matrix}
    \right]
  S^{-1} \\
  &= S
  \left[
    \begin{matrix}
      A & B\\
      C & D
    \end{matrix}
    \right]
  S^{-1}
  = \left[
    \begin{matrix}
      A & -B\\
      -C & D
    \end{matrix}
    \right]
  \end{aligned}
  \end{equation}
  Therefore, flipping the vertical dispersion doesn't change the sign or
  value of coupled Ripken Twiss functions, and neither does the off-energy
  optics dependence on sextupole strength
  $\frac{1}{\partial{K_2}}\frac{\partial{\beta}}{\partial\delta}$. This
  property has been numerically confirmed with the \textsc{mad-x} and
  \textsc{elegant} computations. It can also be used to validate online
  measurements.

\bibliography{oeo.bib}

\end{document}